\documentclass[twoside,twocolumn,10pt]{article}
\usepackage[utf8x]{inputenc}

\usepackage[widespace,poorman]{fourier} 
\usepackage{microtype}

\usepackage{graphicx}
\usepackage{subfigure}
\usepackage{float}
\usepackage{amsmath}
\usepackage{amssymb}
\usepackage{tikz}
\usepackage{pgfplots}
\usepackage{enumerate}
\usepackage{fancybox}
\usetikzlibrary{arrows,snakes,shapes}
\usepackage{epstopdf}
\usepackage{hyperref}
\usepackage{cases}
\usepackage{array}
\usepackage{multirow}
\usepackage{booktabs}
\usepackage {setspace}
\usepackage{appendix}
\usepackage{geometry}
\usepackage[T1]{fontenc}

\usepackage{fancybox}
\usepackage{fancyhdr}
\hypersetup{
backref=true,            
pagebackref=true,    
hyperindex=true,      
colorlinks=true,        
breaklinks=true,       
urlcolor= orange,        
linkcolor= blue,        
citecolor=red,
bookmarks=true,       
bookmarksopen=false,  
}

\newcommand{\sgn}{\mathop{\mathrm{sgn}}}

\usepackage[font=small,labelfont={bf,sf}]{caption}
\usepackage{paralist}
\usepackage{multicol}

\usepackage{abstract}
	
\usepackage{titlesec}
\titleformat{\section}[block]{\center\large\scshape\bfseries}{\thesection.}{0.6em}{}
\titleformat{\subsection}[block]{\normalsize\bfseries}{\thesubsection}{0.6em}{}
\titleformat{\subsubsection}[block]{\normalsize\itshape}{\thesubsubsection}{0.6em}{}

\fancyheadoffset[LE,RO]{0cm}

\fancypagestyle{plain}{%

\fancyhf{} 
}

\title{\vspace{-15mm}%
	\sffamily
	\textbf{Prediction of the dynamic oscillation threshold in a clarinet model with a linearly increasing blowing pressure}
	}	
\author{%
	\large
	\textsc{B. Bergeot$^{a,}$\footnote{Corresponding author, \texttt{baptiste.bergeot@univ-lemans.fr}} , A. Almeida$^{a}$, C. Vergez$^{b}$, B. Gazengel$^{a}$}}

\date{{\small \textit{$^{a}$LUNAM Universit\'{e}, Universit\'{e} du Maine, UMR CNRS 6613, Laboratoire d’Acoustique, Avenue Olivier Messiaen, 72085 Le Mans Cedex 9, France}}\\
{\small \textit{$^{b}$Laboratoire de M\'{e}canique et Acoustique (LMA, CNRS UPR7051), 31 Chemin Joseph Aiguier, 13402 Marseille Cedex 20, France}}}

\begin{document}

\twocolumn[
    \maketitle
    \hrulefill
\begin{onecolabstract}
\noindent Reed instruments are modeled as self-\sloppy sustained oscillators driven by the pressure inside the mouth of the musician. A set of nonlinear equations connects the control parameters (mouth pressure, lip force) to the system output, hereby considered as the mouthpiece pressure. Clarinets can then be studied as dynamical systems, their steady behavior being dictated uniquely by the values of the control parameters. Considering the resonator as a lossless straight cylinder is a dramatic yet common simplification that allows for simulations using nonlinear iterative maps.

This paper investigates analytically the effect of a linearly increasing blowing pressure on the behavior of this simplified clarinet model. When the control parameter varies, results from the so-called \textit{dynamic} bifurcation theory are required to properly analyze the system. This study highlights the phenomenon of \textit{bifurcation delay} and defines a new quantity, the \textit{dynamic oscillation threshold}. A theoretical estimation of the \textit{dynamic oscillation threshold} is proposed and compared with numerical simulations.

\end{onecolabstract}
   \hrulefill
\vspace{0.7cm}]
{
  \renewcommand{\thefootnote}%
    {\fnsymbol{footnote}}
  \footnotetext[1]{{Corresponding author, \texttt{baptiste.bergeot@univ-lemans.fr}}
}

\section{Introduction}
\label{sec:introduction}

One of the interests of mathematical models of musical instruments is to be able to predict certain characteristics of the produced sound given the gesture performed by the musician. In the case of a clarinet for instance, the amplitude, frequency or spectral content (the sound parameters) can be to a certain extent, determined as a function of the blowing pressure and lip force applied to the reed (the control parameters). A basic model, such as the one introduced by Wilson and Beavers \cite{WilJASA1974}, allows to compute the amplitude of the oscillating resonator pressure from the knowledge of these two control parameters, giving results that follow the major tendencies observed in experiments. Several degrees of refinement can be added to this model, usually aiming at realistic sound and mechanical behavior. Well known simplifications of this model allow to study analytically the behavior of the clarinet. Simplified models, of course, are unable to describe or predict with refinement the exact harmonic content of the sound, or the influences of such important details as the reed geometry and composition or the vocal tract of the player. However, they can provide an understanding of the factors essential for the production of sound.

The highest degree of simplification of the model (introduced in Section \ref{sec:elementary-model}) considers a straight, lossless (or losses independent of frequency) resonator and the reed as an ideal spring \cite{Maga1986,MechOfMusInst,Cha08Belin}. With these assumptions, the system can be simply described by an iterated map~\cite{McIntyre83:JASA}. Iterated maps often describe a succession of different regimes with variable periodicity. By analyzing the asymptotic values of these regimes it is possible to estimate: thresholds of oscillation, extinction, beating regimes, etc. \cite{dalmont:3294}, amplitudes and stability of the steady state regime \cite{OllivActAc2005} and phenomena of period doubling \cite{KergoCFA2004,NonLin_Tail_2010}.

These characteristics arise from the so-called \textit{static} bifurcation theory assuming that control parameters are constant. For example, these studies allow to find a \textit{static} oscillation threshold $\gamma_{st}$ \cite{dalmont:3294} such that a constant regime is stable if the blowing pressure is below $\gamma_{st}$ and a periodic regime is stable if it is above $\gamma_{st}$. More precisely, the oscillation emerges through a flip bifurcation \cite{KuzAppBif2004}. This behavior is static, obtained by choosing a constant blowing pressure, letting the system reach its final state, and repeating the procedure for other constant blowing pressures. Therefore, most studies using iterated map approach are restricted to a steady state analysis of the oscillation, even if transients are studied. They focus on the asymptotic amplitude regardless of the history of the system.

During a note attack transient the musician varies the pressure in her/his mouth before reaching a quasi-constant value. During this transient the blowing pressure cannot be regarded as constant. In a mathematical point of view increasing the control parameter (here the blowing pressure) makes the system non-autonomous and results from \textit{static} bifurcation theory are not sufficient to describe its evolution. Indeed, it is known that, when the control parameter varies, the bifurcation point -- i.e. the value of the blowing pressure where the system begins to oscillate -- can be considerably delayed \cite{Kapral1985,Baesens1991,Fruchard2007}. Indeed, the bifurcation point is shifted from $\gamma_{st}$ to a larger value $\gamma_{dt}$ called \textit{dynamic} oscillation threshold. This phenomenon called \textit{bifurcation delay} is not predicted by the \textit{static} theory. Therefore, when the control parameter varies, results from the so-called \textit{dynamic} bifurcation theory are required to properly analyze the system.

The purpose of this paper is to use results from \textit{dynamic} bifurcation theory to describe analytically a simplified clarinet model taking into account a blowing pressure that varies linearly with time. In particular we propose a theoretical estimation of the dynamic oscillation threshold. 

Section \ref{sec:elementary-model} introduces the simplified mathematical model of a clarinet and the iterated map method used  to estimate the existence of the oscillations inside the bore of the clarinet. Some results related to the steady state are presented in this section. Section \ref{sec:model} is devoted to the study of the dynamic system that takes into account a linearly increasing blowing pressure. The phenomenon of bifurcation delay is demonstrated using numerical simulations. A theoretical estimation of the \textit{dynamic oscillation threshold} is also presented and compared with numerical simulations. In Section \ref{sec:Inf_Par} the limits of this approach are discussed. It is shown, when the model is simulated, that the precision (the number of decimal digits used by the computer) has a dramatic influence on the bifurcation delay. The influence of the speed at which the blowing pressure is swept is also discussed. 

\section {State of the art}
\label{sec:elementary-model}

\subsection{Elementary model}

The model of the clarinet system used in this article follows an extreme simplification of the instrument, which can be found in other theoretical works \cite{Maga1986,Cha08Belin}. 

This basic model separates the instrument into two functional elements. One of these is the bore, or resonator, a linear element where the pressure waves propagate without losses. The other is the reed-mouthpiece system, which is considered as a valve controlled by the pressure difference between the mouth and the mouthpiece. It is often called the generator and is the only nonlinear part of the instrument. A table of notation is provided in \ref{App:A}.
 
\begin{figure}[t]
\centering
\includegraphics[width=75mm,keepaspectratio=true]{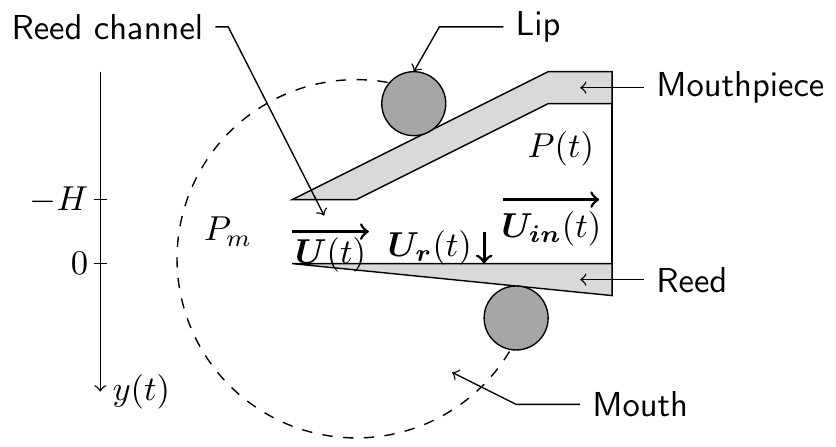}
\caption{Schematic diagram of a single-reed mouthpiece. Presentation of variables, control parameters and choice of axis orientation. $U$ is the flow created by the pressure imbalance $P_m-P$  between the mouth and the bore, $U_r$ is the flow created by the motion of the reed, $U_{in}$ is the flow entering the instrument, $y$ represents the position of the tip of the reed and H is the opening of the reed channel at rest.}
\label{schema_bec_bec}
\end{figure}

\subsubsection{The reed-mouthpiece system}

The reed-mouthpiece system is depicted in Fig. \ref{schema_bec_bec}. The reed is assumed to behave as an ideal spring characterized by its static stiffness per unit area $K_s$. So, its response $y$ to the pressure difference $\Delta P=P_m-P$ is linear and is given by:

\begin{equation}
y=-\frac{\Delta P}{K_s}.
\label{eq_ideal_reed}
\end{equation}

From (\ref{eq_ideal_reed}) we can define the \emph{static closing pressure} $P_M$ which corresponds to the lowest pressure that completely closes the reed channel ($y=-H$):

\begin{equation}
P_M=K_sH.
\label{close_press}
\end{equation}

The reed model also considers that the flow created by the motion of the reed $U_r$ is equal to zero, so that the only flow entering the instrument is created by the pressure imbalance between the mouth and the bore:

\begin{equation}
U_{in}=U.
\end{equation} 

The non-linearity of the reed-mouthpiece system is introduced by the Bernoulli equation which relates the flow $U$ to the acoustic pressure $P$ \cite{HirschAal1990,MOMIchap7}. This relation is the nonlinear characteristics of the exciter, given by:

\begin{subnumcases}{\label{nonlin_carac_2eq}U=}
U_A\left(1-\frac{\Delta P}{P_M}\right)\sqrt{\frac{|\Delta P|}{P_M}}\sgn(\Delta P) \nonumber \\
 \hspace{1.5cm} \text{if} \ \Delta P <P_M \ ; \\
0 \nonumber \\
 \hspace{1.5cm}  \text{if} \ \Delta P >P_M. 
\end{subnumcases}

The flow $U_A$ is calculated using the Bernoulli theorem:
\begin{equation}
U_A=S\sqrt{\frac{2P_M}{\rho}},
\end{equation}
where $S$ is the opening cross section of the reed channel at rest and $\rho$ the density of the air.

Introducing the dimensionless variables and control parameters \cite{Cha08Belin}:
\begin{equation}
\begin{array}{ll}
\Delta p&= {\Delta P}/{P_M}\\
p&={P}/{P_M}\\
u&={Z_c \,U}/{P_M}\\
\gamma &= {P_m}/{P_M} \\
\zeta &={Z_c \,U_A}/{P_M}.
\end{array}
\label{eq:adim}
\end{equation}
$Z_c=\rho c/S_{res}$ is the characteristic impedance of the cylindrical resonator of cross-section $S_{res}$ ($c$ is the sound velocity). Equation (\ref{nonlin_carac_2eq}) becomes:

\begin{subnumcases}{\label{nonlin_carac_2eq_ad}F(p)=}
\zeta \left(1-\gamma +p \right)\sqrt{|\gamma -p|}\sgn(\gamma-p) \nonumber \\  
\hspace{1.5cm} \text{if} \ \gamma -p <1 \ ; \\
0 \nonumber\\
\hspace{1.5cm} \text{if} \ \gamma -p > 1.\label{carNL_beatreed}
\end{subnumcases} 

The parameters $\gamma$ and $\zeta$ are the control parameters of the system. An example of the function $F$ is shown in Fig. \ref{rep_F}.

\subsubsection{The resonator}

Assuming that only plane waves exist in the resonator and propagate linearly, the resonator can be characterized by its reflection function $r(t)$. The general expression relating $p(t)$ to $u(t)$ through $r(t)$ is:

\begin{equation}
p(t)-u(t)=[r\ast (p+u)](t).
\label{eq_resonator}
\end{equation}

The resonator is modeled as a straight cylinder. Reflections at the open end of the resonator are considered perfect (no radiation losses) and viscous and thermal losses are ignored. In this case the reflection function becomes a simple delay with sign inversion:

\begin{equation}
r(t)=-\delta(t-\tau),
\label{lossless_r}
\end{equation}
where $\delta$ is the Dirac generalized function and $\tau=2l/c$ is the round trip time of the sound wave with velocity $c$ along the resonator of length $l$.
 
With the reflection function (\ref{lossless_r}), equation (\ref{eq_resonator}) becomes:
\begin{equation}
p(t)-u(t)=-\left[p(t-\tau)+u(t-\tau)\right].
\end{equation}

Assuming that the blowing pressure $\gamma$ skips instantaneously from 0 to a finite value and remains constant, $p$ and $u$ remain constant during the first half-period and hence during each forthcoming half-period. Therefore, $p$ and $u$ are square waves.

Using a discrete time formulation (the discretization is done at regular intervals $\tau$) and noting $p(n\tau)=p_n$ and $u(n\tau)=u_n$, we obtain the following difference equation:
\begin{equation}
p_n-u_n=-\left(p_{n-1}+u_{n-1}\right).
\label{lossless_r_discr}
\end{equation}

\begin{figure*}[t]
\centering
\subfigure[function $F$]{\includegraphics[width=75mm,keepaspectratio=true]{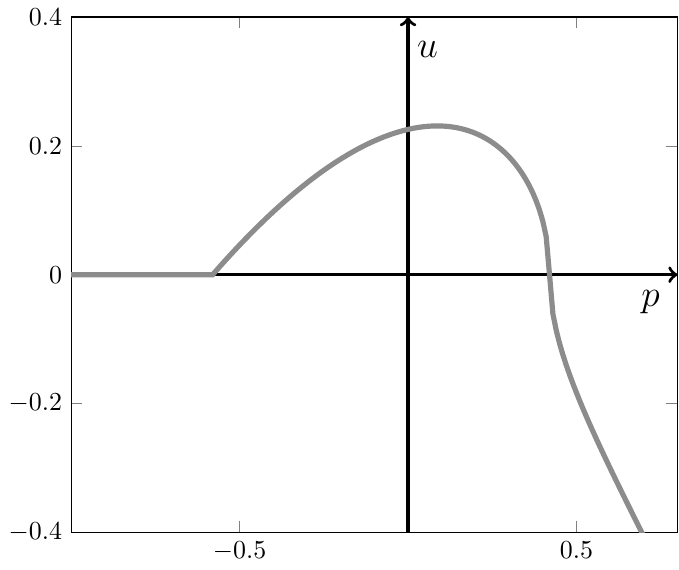}\label{rep_F}}
\subfigure[function $G$]{\includegraphics[width=75mm,keepaspectratio=true]{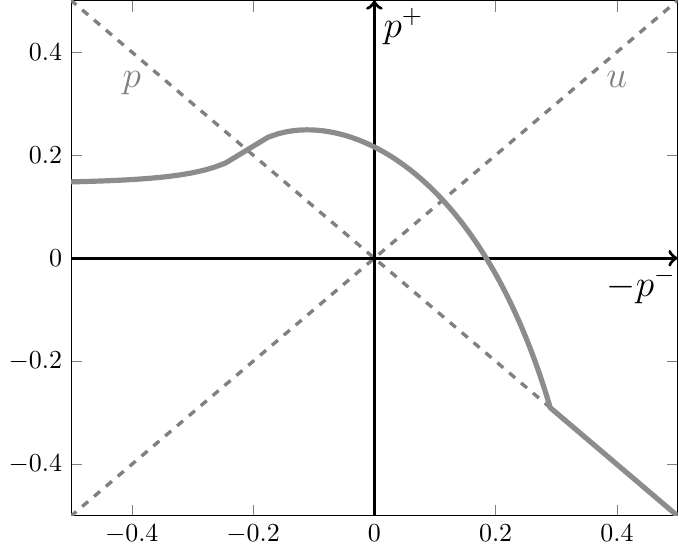}\label{rep_G}}
\caption{Nonlinear characteristics in $u=F(p)$ representation (a) and $p^+=G(-p⁻ )$ representation (b) for $\gamma=0.42$ and $\zeta=0.6$.}   
\end{figure*}

\subsection{Iterated map: outgoing and incoming wave representation}

In linear acoustics any planar wave can be expanded into an outgoing wave $p^+$ and an incoming wave $p^-$. Using the dimensionless variables defined in equation \eqref{eq:adim}, the acoustic pressure $p$ and flow $u$ are given by:

\begin{align}
p&= p^+ +p^- & ; & &u &=p^+ -p^-,
\label{pu_to_pppm}
\end{align}

Replacing in equation~\eqref{lossless_r_discr},
\begin{align}
p^+&= \frac{1}{2}(p+u) & ; & &p^- &=\frac{1}{2}(p-u).
\end{align}

By combining equations (\ref{nonlin_carac_2eq_ad}) and (\ref{pu_to_pppm}) a nonlinear relation $G$ between $p^+$ and $p^-$ can be obtained:
\begin{equation}
p^+=G\left(-p^-\right).
\label{NonLinRelpppm}
\end{equation}

An explicit expression of the function $G$ was determined, for $\zeta<1$, by Taillard et al. \cite{NonLin_Tail_2010}. Fig. \ref{rep_G} shows an example of the function $G$. Using equations (\ref{pu_to_pppm}), the relation (\ref{lossless_r_discr}) becomes:
\begin{equation}
p^-_n=-p^+_{n-1}.
\label{ResEqpppm}
\end{equation}

Finally, equations (\ref{NonLinRelpppm}) and (\ref{ResEqpppm}) define the iterated map \cite{Maga1986,McIntyre83:JASA}:

\begin{equation}
p^+_n=G\left(-p^-_n\right)=G\left(p^+_{n-1}\right).
\label{DiffEq_G}
\end{equation}

In the following, the variable $p^+$ will be used preferentially. The variable $p$ can easily be calculated using equations (\ref{ResEqpppm}) and  (\ref{pu_to_pppm}).

\subsection{Results from static bifurcation theory} 

The difference equation (\ref{DiffEq_G}) can be analyzed using the \textit{static} bifurcation theory, which assumes that the control parameters are constant. This will be hereafter referred to as the \textit{static case}. The parameter $\gamma$ will be specifically introduced as a subscript in the definition of the nonlinear characteristics (\ref{DiffEq_G}), stressing that this will be the parameter of interest in the current study ($\zeta$ will always consider to be constant): 
\begin{equation}
p^+_n=G_{\gamma}\left(p^+_{n-1}\right).
\label{DiffEq_G_InflGamm}
\end{equation}

Some of the predictions of the \textit{static} bifurcation theory that are important to this work are recalled in the following sections while applying them to the map of equation (\ref{DiffEq_G_InflGamm}) \cite{MechOfMusInst,Cha08Belin,NonLin_Tail_2010}.

\subsubsection{Expression of the static regime and static oscillation threshold}

For all values of the control parameter $\gamma$ below a particular value of the parameter $\gamma$ called \textit{static oscillation threshold} and noted $\gamma_{st}$ the series $p^+_n$ converges to a single value (the static regime), also referred to as the fixed point of $G_{\gamma}$. It can be found by solving the following equation:

\begin{equation}
p^{+*}=G_{\gamma}\left(p^{+*}\right).
\label{fixed_point_G_def}
\end{equation}

After solving the equation we obtain :
\begin{equation}
p^{+*}(\gamma)=\frac{\zeta}{2}(1-\gamma)\sqrt{\gamma}.
\label{fixed_point_G}
\end{equation}

When the static regime is reached $p^+_{n}=p^+_{n-1}=-p^-_n$. Therefore, for the variable $p=p^++p^-$, the static regime is equal to zero. 

The static regime exists for all values of the parameter $\gamma$ but it is stable when $\gamma<\gamma_{st}$ and unstable when $\gamma>\gamma_{st}$. The condition of stability of the static regime \cite{MechOfMusInst} allowing to obtain the value of the static oscillation threshold is:

\begin{equation}
\left|G_{\gamma}'\left(p^{+*}\right) \right| <1,
\label{Jacoc}
\end{equation}
where $G_{\gamma}'$ is the first derivative of the function $G_{\gamma}$. The value of the static oscillation threshold is finally: 

\begin{equation}
\gamma_{st} =\frac{1}{3}.
\end{equation}

Beyond the oscillation threshold, other bifurcations occur, the 2-valued oscillating regime becoming unstable and giving rise to a 4-valued oscillating state. This cascade is  the classical scenario of successive period doublings, leading eventually to chaos \cite{Feigen1979,NonLin_Tail_2010}. The values of the parameter $\gamma$ for which appear the different 2n-valued oscillating regimes depend on the value of the parameter $\zeta$: the smaller is $\zeta$, the earlier the  2n-valued oscillating regimes appear. When $\gamma=1/2$, whatever the value of $\zeta$, a 2-valued oscillating regime reappears, the beating-reed regime . This is a particularity of model of the clarinet, it is due to the fact that when $\gamma-p>1$ (equation (\ref{carNL_beatreed})) the reed presses against the mouthpiece lay. It can be shown \cite{Cha08Belin} that in this permanent regime $p=\pm \gamma$ (c.f. Fig. \ref{stat_bif_digram}).

\subsubsection{\textit{Static} bifurcation diagrams}

Common representations of the \textit{static} bifurcation diagram for clarinets usually show the steady state of the pressure inside the mouthpiece $p$ or that of its amplitude (corresponding in the lossless model to the absolute value of $p$) with respect to the control parameter $\gamma$ \cite{dalmont:3294}. In this paper, calculations are based on $p^+$, so that most bifurcation diagrams will represent the steady state of the outgoing wave~\cite{NonLin_Tail_2010}.

Fig. \ref{stat_bif_digram} shows an example of these three representations of the \textit{static} bifurcation diagram for $\zeta=0.5$. Fig.~\ref{stat_bif_digram} represents only the two first branches of the diagrams. The first branch corresponds to the fixed points of the function $G_{\gamma}$ and the second branch represents the fixed points of the function $(G_{\gamma} \circ G_{\gamma})$. On Fig. \ref{diagbif_p} and Fig. \ref{diagbif_pp} the dashed line represents the curve of the static regime. For the variable $p$, the static regime is equal to zero and for the variable $p^+$ it is a function of the parameter $\gamma$, noted $p^{+*}(\gamma)$. Oscillating regimes with higher periodicities which may appear between $\gamma=1/3$ and $\gamma=1/2$ are not represented.

\begin{figure}[h!]
\centering
\subfigure[]{\includegraphics[width=65mm,keepaspectratio=true]{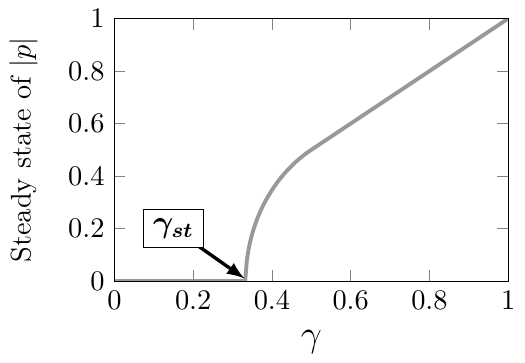}}
\subfigure[]{\includegraphics[width=65mm,keepaspectratio=true]{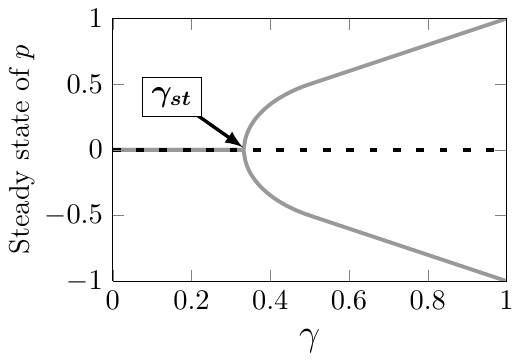}\label{diagbif_p}}
\subfigure[]{\includegraphics[width=65mm,keepaspectratio=true]{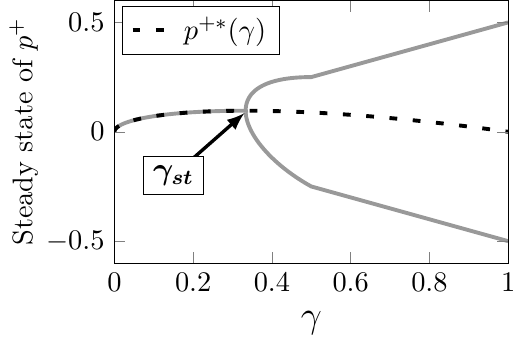}\label{diagbif_pp}}
\caption{Graphical representation of the static bifurcation diagrams for $\zeta=0.5$. Diagrams based on variables (a) $|p|$, (b) $p$ and  (c) $p^+$. The dashed line represents the curve of the static regime, corresponding to the fixed point $p^{+*}(\gamma)$ of the function $G_{\gamma}$ when the diagram is base on variable $p^+$.}
\label{stat_bif_digram}
\end{figure}

\section{Time-varying blowing pressure}
\label{sec:model}

\subsection{Problem statement}

\subsubsection{Definitions}

Before presenting the problem, some definitions are introduced in order to avoid ambiguity in the vocabulary used hereafter. In the remainder of this paper, all simulations and calculations will be performed considering that the parameter $\zeta$ is a constant and equal to 0.5.  The definitions presented below, used commonly in works dealing with bifurcation theory, can present some conflicts with that of musical acoustics. The terms that will be used in the remaining discussions are clarified in the following paragraphs:  

\paragraph{Static case} The control parameter $\gamma$ is constant and the system is described by:

\begin{figure*}[t]
\centering
\subfigure[]{\includegraphics[width=75mm,keepaspectratio=true]{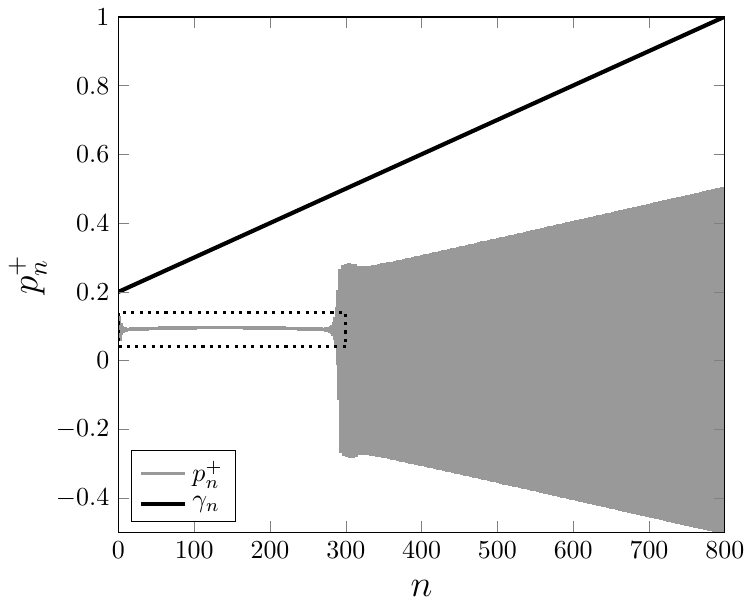}}
\subfigure[]{\includegraphics[width=75mm,keepaspectratio=true]{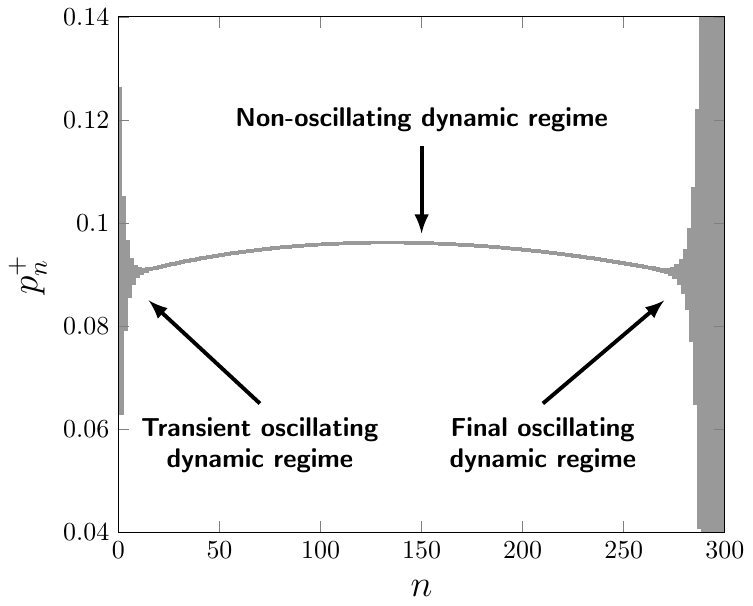}}
\caption{Numerical simulation performed on the system (\ref{dynsys_pp}). (a) complete orbit of the series and (b) zoom near the non-oscillation dynamic regime. $\zeta=0.5$, $\epsilon=10^{-3}$, $\gamma_0=0.2$ and $p^+_0=G(0,\gamma_0)$.}
\label{seuildynVSci_0}
\end{figure*}

\begin{equation}
p^+_n=G_{\gamma}\left(p^+_{n-1}\right).
\label{DiffEq_G_InflGamm_1}
\end{equation}

The steady state of the series $p^+_n$ depends on the value of the control parameter $\gamma$. If $\gamma$ is smaller than $\gamma_{st}$, the series tends to a static regime. To avoid confusion, the static regime will now be called \textit{non-oscillating static regime}. If $\gamma$ is larger than $\gamma_{st}$ the steady state of $p^+_n$ is an oscillating regime. This regime is called \textit{oscillating static regime}. This behavior is still static, obtained by choosing a value of $\gamma$, letting the system reach its steady state, and repeating the procedure for each value of $ \gamma$. Note that, even if the system tends to a steady state, the initial condition $p_{0}^{+}$ often induces a transient regime.

\paragraph{Dynamic case} As pointed in the introduction, in a musical context, the blowing pressure cannot always be considered constant. The dynamic case take this into account considering that the control parameter $\gamma$ is variable and now written as $\gamma_n$. When $\gamma$ is a linear function of time, the system is described by the following difference equations:

\begin{subnumcases}{\label{dynsys_pp}}
p^+_n=G\left(p^+_{n-1},\gamma_n\right)\label{dynsys_pp_a}\\
\gamma_{n}=\gamma _{n-1} +\epsilon.
\end{subnumcases}

Since $\gamma$ is changed only at each multiple of $\tau$, the solution of equation (\ref{dynsys_pp}) is still a square signal, i.e. two-state oscillating regime.

A slowly varying parameter implies that $\epsilon$ is arbitrarily small ($\epsilon\ll 1$). The hypothesis of an arbitrarily small $\epsilon$ could be questioned in the context of the playing of a musical instrument. However, this hypothesis is required in order to use the framework of dynamic bifurcation theory (see forthcoming sections).

An example of a numerical simulation performed on the system (\ref{dynsys_pp}) is shown in Fig.~\ref{seuildynVSci_0} for $\zeta=0.5$, $\epsilon=10^{-3}$ and an initial condition $\gamma_0=0.2$. The initial value of the outgoing wave is $p^+_0=G(-p^-_0=0,\gamma_0)$. Indeed, for $n=0$ the incoming wave $p^-$ is clearly zero, otherwise sound would have traveled back and forth with an infinite velocity.

The series $p^+_n$  first shows a short oscillating transient, which will be called \textit{transient oscillating} \textit{dynamic regime}. This oscillation decays into a \textit{non-oscillating dynamic regime}. Beyond a certain threshold, a new oscillation grows, giving rise to the \textit{final oscillating dynamic regime}.

This paper will focus on the transition (i.e. the bifurcation) from the \textit{non-oscillating dynamic regime} to the \textit{final oscillating dynamic regime}. The value of the parameter $\gamma$ for which the bifurcation occurs is called \textit{dynamic oscillation threshold}, noted $\gamma_{dt}$.

\subsubsection{Bifurcation delay}
 
\textit{Bifurcation delay} occurs in nonlinear-systems with time varying control parameters. Fruchard and
Schäfke \cite{Fruchard2007} published an overview of the problem of bifurcation delay.

In fig.~\ref{BIF_DELAY}, the system (\ref{dynsys_pp}) was simulated numerically, showing the time evolution of the series $p^+_n$ and of the control parameter $\gamma_n$ (cf. Fig. \ref{BIF_DELAYa}). To better understand the consequence of a time-varying parameter, the orbit of the series $p^+_n$ is plotted as a function of the parameter $\gamma_n$ -- in this case the evolution of the system can be interpreted as a \textit{dynamic bifurcation diagram}. This  is compared to the static bifurcation diagram in Fig. \ref{BIF_DELAYb}. We can observe that the static and the dynamic bifurcation diagrams coincide far from the static oscillation threshold $\gamma_{st}$. However, in the dynamic case, we can see that the orbit continues to follow closely the branch of the fixed point of function $G$ throughout a remarkable extent of its unstable range, i.e. after $\gamma_{st}$: the bifurcation point is shifted from  the static oscillation threshold $\gamma_{st}$ to the dynamic oscillation threshold~$\gamma_{dt}$. The term \emph{bifurcation delay} is used to state the fact that the static oscillation threshold $\gamma_{st}$ is smaller than the dynamic oscillation threshold $\gamma_{dt}$.

\begin{figure*}[t]
\centering
\subfigure[]{\includegraphics[width=75mm,keepaspectratio=true]{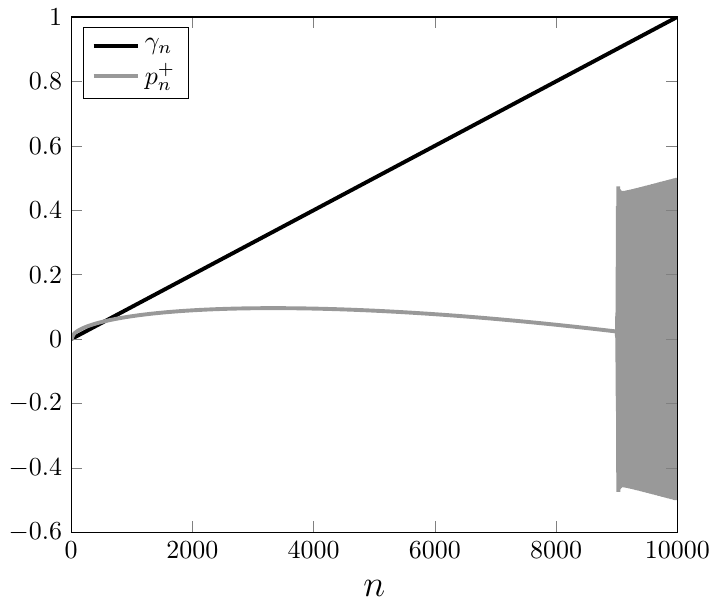}\label{BIF_DELAYa}}
\subfigure[]{\includegraphics[width=75mm,keepaspectratio=true]{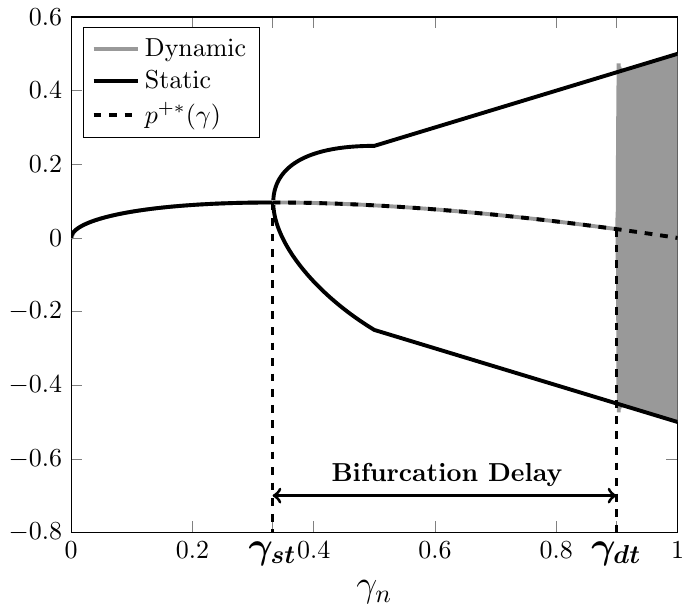}\label{BIF_DELAYb}}
\caption{(a) Time evolution of the series $p^+_n$ and of the control parameter $\gamma_n$.(b) Comparison between the series $p^+_n$ and the \textit{static} bifurcation diagram as a function of $\gamma_n$. $\zeta=0.5$, $\epsilon=10^{-4}$, $\gamma_0=0$ and $p^+_0=G(0,\gamma_0)$.}
\label{BIF_DELAY}
\end{figure*}

Non-standard analysis has been used in the past to study the phenomenon of bifurcation delay \cite{FruchInstFour1996,FruchaScaf2003}, explaing that one of the causes of the bifurcation delay is the exponential proximity between the orbit of the series $p^+_n$ and the curve of the the fixed point of $G$. Other studies of bifurcation delay using standard mathematical tools -- mathematics \cite{Baesens1991,Baesens1995} or physics publications \cite{Kapral1985,Tre_AmJPhy_2004} -- explain bifurcation delay as an accumulation of stability during the range of $\gamma$ for which the fixed point of $G$ is stable (i.e. $0<\gamma_n<\gamma_{st}$). The dynamic oscillation threshold therefore appears as the value of the parameter $\gamma$ at which the stability previously accumulated is compensated.

In musical acoustics literature some papers present results showing the phenomenon of bifurcation delay without never making a connection to the concept of \textit{dynamic bifurcation}. For example this phenomenon is observed  in simulations of clarinet-like systems using a slightly more sophisticated clarinet model (Raman's model) \cite{AtDalGil_ApplAcous2004}. Raman's model takes losses into account although they are assumed to be independent of frequency (see \cite{dalmont:3294} for further explanation). Bifurcation delay can also explain the difficulty in estimating the static oscillation threshold by using a slowly variable blowing pressure \cite{DiFeCFA2010}. In a preliminary work \cite{BBCFA2012Exp}, bifurcation delays were experimentally observed in a clarinet-like instrument.

\subsection{Analytical study of the dynamic case}

\begin{figure*}[t]
\centering
\subfigure[$\epsilon=10^{-2}$]{\includegraphics[width=75mm,keepaspectratio=true]{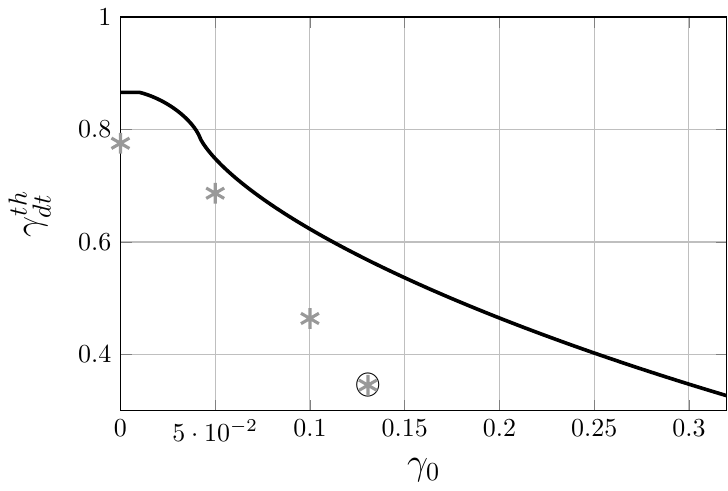}}
\subfigure[$\epsilon=10^{-3}$]{\includegraphics[width=75mm,keepaspectratio=true]{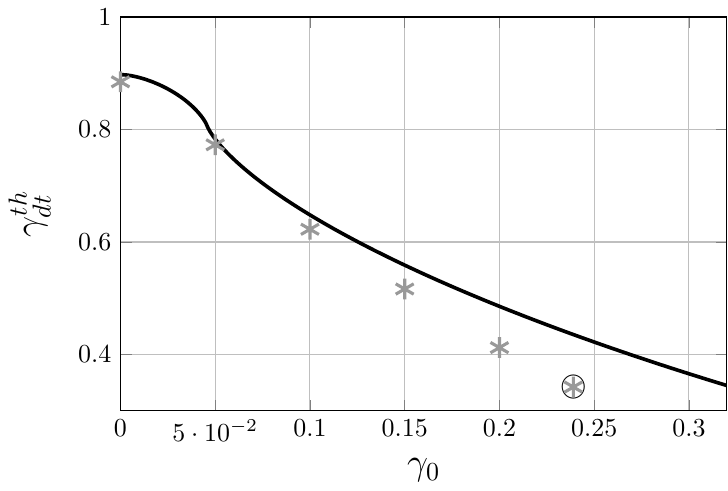}}
\subfigure[$\epsilon=10^{-4}$]{\includegraphics[width=75mm,keepaspectratio=true]{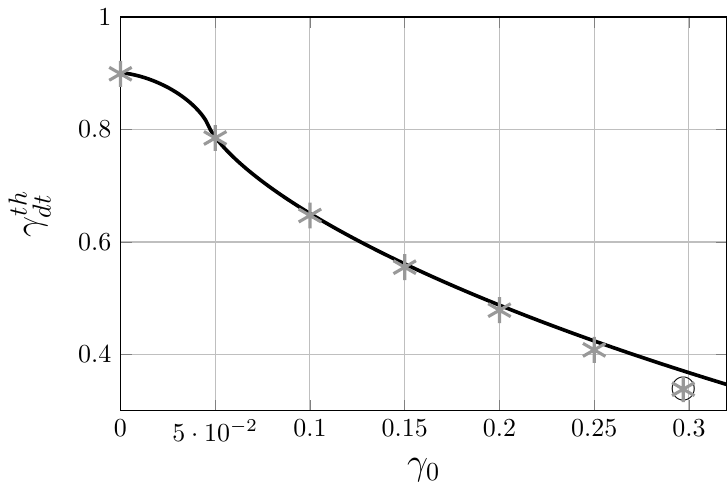}}
\subfigure[]{\includegraphics[width=75mm,keepaspectratio=true]{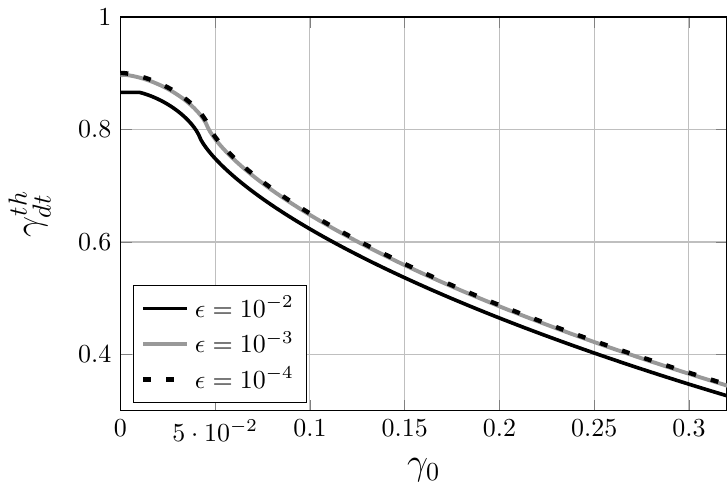}\label{seuildynVSci_d}}
\caption{Plot of $\gamma_{dt}$ as a function of the initial condition $\gamma_0$, for different values of the slope $\epsilon$. (a), (b) and (c): solid lines are the theoretical prediction $\gamma_{dt}^{th}$ calculated from equation (\ref{dynoscthre_2}). Gray $"\ast"$ markers represent the value $\gamma_{dt}^{num}$ for which the system begins to oscillate. (d): combination of the previous theoretical predictions. $"\circledast"$ represent the highest $\gamma_0$ for which the system has enough time to reach a non-oscillating dynamic regime.}
\label{seuildynVSci}
\end{figure*}

This section presents an analytical description of a clarinet-like system in a dynamic case. The notion of \textit{invariant curve} ($\phi(\gamma,\epsilon)$), invariant under the mapping (\ref{dynsys_pp}), will be needed for this study. The study of the stability of the invariant curve allows to define an analytical estimation of the dynamic oscillation threshold. A generic method to calculate the invariant curve is given by Baesens \cite{Baesens1991}\footnote{In \cite{Baesens1991}, the invariant curve is called \textit{adiabatic invariant manifold}.}, based on a perturbation method \cite{Book_bender_Orszag}.

\subsubsection{Invariant curve}

The invariant curve $\phi(\gamma,\epsilon)$ is invariant under the mapping (\ref{dynsys_pp}), satisfying the following equation:

\begin{equation}
\phi(\gamma,\epsilon)=G\left(\phi(\gamma-\epsilon,\epsilon),\gamma\right).
\label{eqdiff_1}
\end{equation}

This curve plays a similar role for the dynamic system as fixed points for the static system, attracting or repelling the orbits. It is independent of the initial condition.

First of all, the invariant curve is expanded into a power series of $\epsilon$, here truncated to the first order:

\begin{equation}
\phi(\gamma,\epsilon) \approx \phi_0(\gamma)+\epsilon \phi_1(\gamma).
\end{equation}

Fig.~\ref{BIF_DELAY} shows that, during the dynamic phase, the orbit of the series $p^+_n$ closely follows the curve of  the fixed points of $G$. This allows to linearize function $G$ around the curve  of the fixed points $p^{+*}(\gamma)$:

\begin{multline}
G(x,\gamma) \approx G\left(p^{+*}(\gamma),\gamma\right)+\\\\ \left[x-p^{+*}(\gamma)\right]\partial_x G\left(p^{+*}(\gamma),\gamma\right),
\end{multline}
using the notation
\begin{equation}
\partial_x G\left(x,y\right)=\frac{\partial G(x,y)}{\partial x},
\end{equation}
and knowing that $G\left(p^{+*}(\gamma),\gamma\right)=p^{+*}(\gamma)$ (cf. equation (\ref{fixed_point_G_def})). Finally, using a Taylor expansion of $\phi(\gamma-\epsilon,\epsilon)$ equation (\ref{eqdiff_1}) is successively solved for the functions $\phi_0(\gamma)$ and $\phi_1(\gamma)$, yielding:

\begin{multline}
\phi(\gamma,\epsilon)\approx p^{+*}(\gamma) \, + \\\\ \epsilon\;\frac{d p^{+*}(\gamma)}{d \gamma}\;
\frac{\partial_xG\left(p^{+*}(\gamma),\gamma\right)}{\partial_xG\left(p^{+*}(\gamma),\gamma\right)-1}.
\label{ppchap_beforeN}
\end{multline}

Using the explicit expressions of $p^{+*}$ and ${d p^{+*}}/{d \gamma}$ we have:

\begin{multline}
\phi(\gamma,\epsilon)\approx \frac{\zeta}{2}(1-\gamma)\sqrt{\gamma} \;  -\\\\  \epsilon\;\frac{\zeta \left( 3\,\gamma-1\right)}{4\,\sqrt{\gamma}}\;
\frac{\partial_xG\left(p^{+*}(\gamma),\gamma\right)}{\partial_xG\left(p^{+*}(\gamma),\gamma\right)-1}.
\label{ppchap_beforeN}
\end{multline}

More details about the calculation of the invariant curve are given in Appendix \ref{an:incurve}.

To simplify the notation, in the rest of the document the invariant curve will be noted $\phi(\gamma)$. Its dependency on parameter $\epsilon$ is not explicitly stated.

\subsubsection{Stability of the invariant curve and theoretical estimation of the dynamic oscillation threshold}

A theoretical estimation of the dynamic oscillation threshold is done by identifying the value of $\gamma$ for which the invariant curve looses its stability. The invariant curve is said to be unstable when the orbit of of the series $p^+_n$ escapes from the neighborhood of the invariant curve $\phi(\gamma,\epsilon)$.

To investigate the stability of the invariant curve $\phi(\gamma,\epsilon)$, the function $G$ in equation (\ref{dynsys_pp_a}) is expanded in a first-order Taylor series around the invariant curve \cite{Baesens1991}:

\begin{multline}
p^+_n = G(p^+_{n-1},\gamma_n)\\\\
\approx G\left(\phi(\gamma_n-\epsilon),\gamma_n\right)+\\\\ \left[p^+_{n-1}-\phi(\gamma_n-\epsilon)\right] \partial_x G\left(\phi(\gamma_n-\epsilon),\gamma_n\right)\label{DL_p_u}.
\end{multline}

A new variable is defined that describes the distance between the actual orbit and the invariant curve:

\begin{equation}
w_n=p^+_n-\phi(\gamma_n),
\end{equation}
and using equation (\ref{eqdiff_1}), equation (\ref{DL_p_u})  becomes:

\begin{equation}
w_n=w_{n-1} \partial_xG\left(\phi(\gamma_n-\epsilon),\gamma_n\right).
\label{series_w}
\end{equation}

The solution of equation (\ref{series_w}) is formally:

\begin{equation}
w_n=w_0\prod_{i=1}^{n} \partial_xG\left(\phi(\gamma_i-\epsilon),\gamma_i\right),
\label{exact_solution_w1}
\end{equation}
for $n \geq 1$ and where $w_0$ is the initial value of $w_n$. The absolute value of $w_n$ can be written as follow:

\begin{multline}
|w_n|=\\|w_0|\exp\left(\sum_{i=1}^{n}\ln\left| \partial_xG\left(\phi(\gamma_i-\epsilon),\gamma_i\right)\right|\right).
\label{exact_solution_w2}
\end{multline}

Finally, using Euler's approximation the sum is replaced by an integral:

\begin{multline}
|w_n|\approx \\ |w_0|\exp\left(\int_{\gamma_0+\epsilon}^{\gamma_n+\epsilon}\ln\left| \partial_xG\left(\phi(\gamma'-\epsilon),\gamma'\right)\right|\frac{d\gamma'}{\epsilon}\right).
\label{exact_solution_w3}
\end{multline}

Equation (\ref{exact_solution_w3}) shows that the variable $p^+$ starts to diverge from the invariant curve $\phi(\gamma,\epsilon)$ when the argument of the exponential function changes from negative to positive. Therefore, the analytical estimation of the dynamic oscillation threshold $\gamma_{dt}^{th}$ is defined by:

\begin{equation}
\int_{\gamma_0+\epsilon}^{\gamma_{dt}^{th}+\epsilon}\ln\left| \partial_xG\left(\phi(\gamma'-\epsilon),\gamma'\right)\right|d\gamma'=0,
\label{dynoscthre_2}
\end{equation}
where $\gamma_0$ is the initial value of $\gamma$. This result can be deduced from \cite{Baesens1991} (equation (2.18)), it may also be obtained in the framework of non-standard analysis \cite{FruchInstFour1992}.

The theoretical estimation $\gamma_{dt}^{th}$ of the dynamic oscillation threshold depends on the initial condition $\gamma_0$ and on the increase rate $\epsilon$, it is therefore written $\gamma_{dt}^{th}(\gamma_0,\epsilon)$. 

A numerical solution $\gamma_{dt}^{th}(\gamma_0,\epsilon)$ of the implicit equation (\ref{dynoscthre_2}) is plotted in Fig. \ref{seuildynVSci} as a function of the initial condition $\gamma_0$ and for $\epsilon=10^{-2}$, $10^{-3}$ and $10^{-4}$. $\gamma_{dt}^{th}$ can be much larger than static oscillation threshold $\gamma_{st}=1/3$ for small initial conditions $\gamma_{0}$. When the initial condition value $\gamma_0$ increases, $\gamma_{dt}^{th}$ approaches the static threshold. Fig. \ref{seuildynVSci_d} shows that the bifurcation delay seems to be independent of the increase rate $\epsilon$ if this value is sufficiently small (typically $\leq 10^{-3}$).

Equation (\ref{dynoscthre_2}) states that when $\gamma=\gamma_{dt}^{th}$ we have $|w_n|\approx|w_0|$, providing a good estimation of the dynamic oscillation threshold $\gamma_{dt}$ if $|w_0|$ is sufficiently small, i.e. if $p^+_0$ is sufficiently close to $\phi(\gamma_0)$. $\gamma_0 = 0$ can be problematic since  $\phi(0,\epsilon)=-\infty$, but a single iteration is sufficient to bring the orbit to a neighborhood of the invariant curve. Therefore, we make the assumption that

\begin{equation}
\gamma_{dt}^{th}(0,\epsilon) \approx \gamma_{dt}^{th}(\epsilon,\epsilon).
\label{assump}
\end{equation}

A non-exhaustive study done by running a few simulations shows that for $\epsilon = 10^{-4}$ the error in $\gamma_{dt}$ due to this approximation is under $10^{-8}$, rising to $10^{-7}$ when $\epsilon = 10^{-3}$ and $2\times10^{-5}$ when $\epsilon = 10^{-2}$.

\begin{figure}[t]
\centering
\includegraphics[width=75mm,keepaspectratio=true]{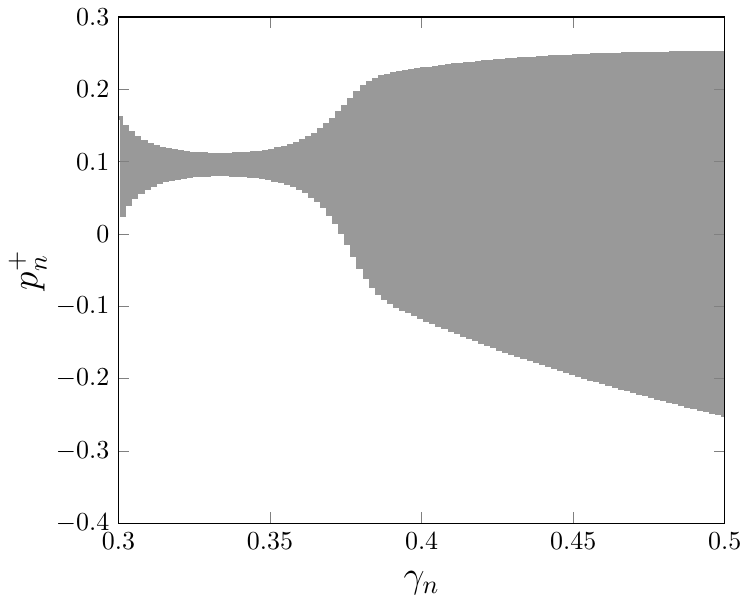}
\caption{Representation of the series $p^+_n$ as a function of $\gamma_n$ for $\zeta=0.5$,$\epsilon=10^{-3}$, $\gamma_0 = 0.3$ and $p^+_0=G(0,\gamma_0)$.}
\label{orbitci03eps1em3}
\end{figure}

\subsection{Benchmark of theoretical estimators for the dynamic threshold}
\label{sec:benchmark}

\begin{figure}[t]
\centering
\subfigure[precision = 5000]{\includegraphics[width=75mm,keepaspectratio=true]{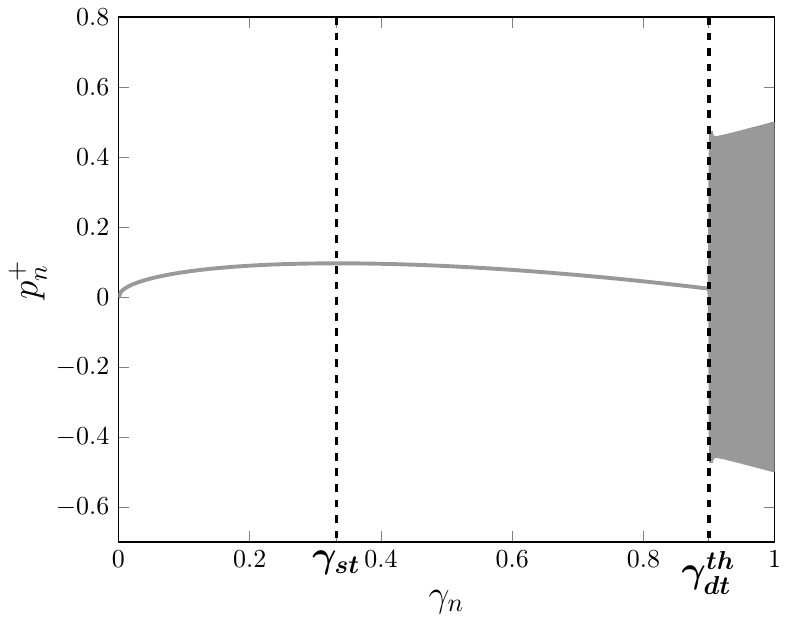}\label{orbitpp_prec5000&15_a}}
\subfigure[precision = 15]{\includegraphics[width=75mm,keepaspectratio=true]{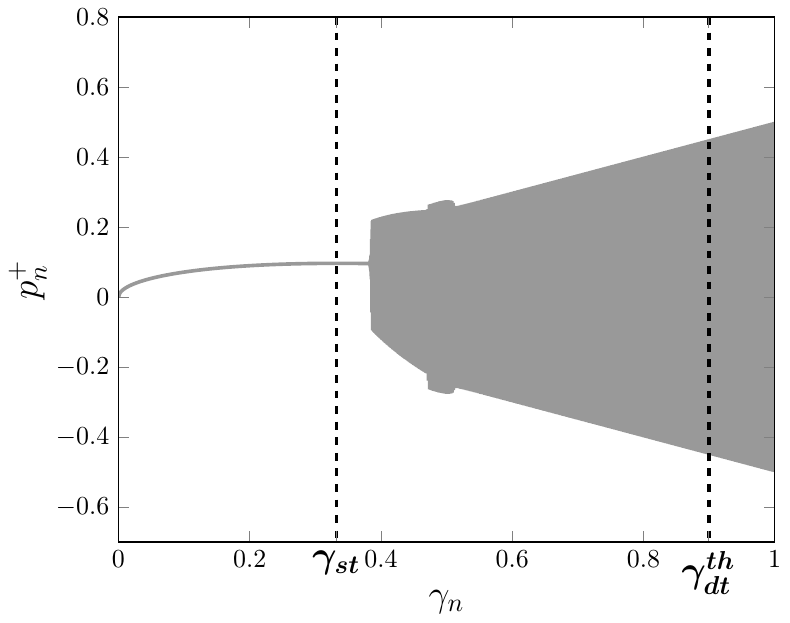}\label{orbitpp_prec5000&15_b}}
\caption{Representation of the series $p^+_n$ for $\zeta=0.5$, $\epsilon=10^{-4}$, $\gamma_0 = 0$, $p^+_0=G(0,\gamma_0)$ and for two different values of the precision.}
\label{orbitpp_prec5000&15}
\end{figure}

Multiple criteria can be associated to the beginning of the oscillating regime. For instance, the oscillations can start before the series departs from the vicinity of the invariant curve as described in equation (\ref{dynoscthre_2}). Moreover, because of the approximation used between equations (\ref{exact_solution_w3}) and (\ref{dynoscthre_2}), the value of $\gamma=\gamma_{dt}^{th}$ may not be an accurate estimation of the value at which the orbit departs from this vicinity. 

For comparison, a dynamic oscillation threshold (noted $\gamma_{dt}^{num}$) is calculated by simulating system (\ref{dynsys_pp}) and compared with $\gamma_{dt}^{th}$. When the orbit of the series $p^+_n$ is periodic, the sign of the second order difference of $p^+_n$ changes sign at each iteration (i.e.~the curve of $p^+_n$ changes from upward to downward concave). In discrete time formulation the second order difference is given by:

\begin{equation}
\left(\delta^2p^+\right)_i=\left(p^+_{i}-p^+_{i-1}\right)-\left(p^+_{i-1}-p^+_{i-2}\right). 
\end{equation}

Therefore, $\gamma_{dt}^{num}$, the oscillation threshold measured in numerical simulations, is reached when
\begin{equation}
\left(\delta^2p^+\right)_{i-1} \, \left(\delta^2p^+\right)_{i} <0,
\end{equation}
is satisfied.

Then, in Fig.~\ref{seuildynVSci}, $\gamma_{dt}^{num}$ is compared with $\gamma_{dt}^{th}$ (gray $"\ast"$ markers). In some cases the series $p^+_n$ never reaches the non-oscillating dynamic regime. An example of such situations is shown in Fig. \ref{orbitci03eps1em3}. The values of $\gamma_{dt}^{num}$ corresponding to the last initial values $\gamma_0$ for which the system has enough time to reach the non-oscillating dynamic regime are circled.

Fig.~\ref{seuildynVSci} shows that for $\epsilon=10^{-4}$ the theoretical result $\gamma_{dt}^{th}$ provides a good estimation of the observed dynamic oscillation threshold. For $\epsilon=10^{-3}$, the theoretical estimation is also good if the the initial condition is sufficiently small but as $\gamma_0$ gets closer to the static threshold $\gamma_{st}$ the system begins to oscillate before $\gamma=\gamma_{dt}^{th}$. Finally, for $\epsilon=10^{-2}$, $\gamma_{dt}^{num}$ is always smaller than $\gamma_{dt}^{th}$.

\section{Limit of the model: influence of the precision}
\label{sec:Inf_Par}

The phenomenon of bifurcation delay is very sensitive to noise: either numerical noise (round-off errors of the computer) or experimental noise (due to turbulence for instance). Indeed, when the static threshold is exceeded the system is very unstable. As a result, to observe bifurcation delay with numerical simulations and compare to theoretical results, it is necessary to perform calculations using a very high precision, as was done previously in this paper. For lower precisions the bifurcation delay can be considerably reduced (see \cite{FruchaScaf2003} for an example in the logistic map). 

Fig.~\ref{orbitpp_prec5000&15} shows an example of numerical simulation performed on system (\ref{dynsys_pp}). Fig. \ref{orbitpp_prec5000&15_a} and Fig. \ref{orbitpp_prec5000&15_b} differ only in the numerical precision (i.e. the number of decimal digits) used to calculate the orbit. The choice of the precision is possible using \emph{mpmath}, the arbitrary precision library of \emph{Python}. Fig.~\ref{orbitpp_prec5000&15_a} was obtained using a precision of 5000 decimal digits, in this case $\gamma_{dt}^{th}$ gives a good estimation of the bifurcation point, as it has already been shown in Fig.~\ref{seuildynVSci}. On the other hand, using a precision of 15 decimal digits (Fig.~\ref{orbitpp_prec5000&15_b}), the bifurcation delay is considerably reduced and the theoretical estimation of the dynamic oscillation threshold is not valid.

To highlight the influence of the precision $\gamma_{dt}^{num}$ is calculated for different precisions. Results are plotted in Fig. \ref{Fig_Influ_prec_1} and compared to the analytical values $\gamma_{st}$ and $\gamma_{dt}^{th}$.

\begin{figure}[t]
\centering
\includegraphics[width=75mm,keepaspectratio=true]{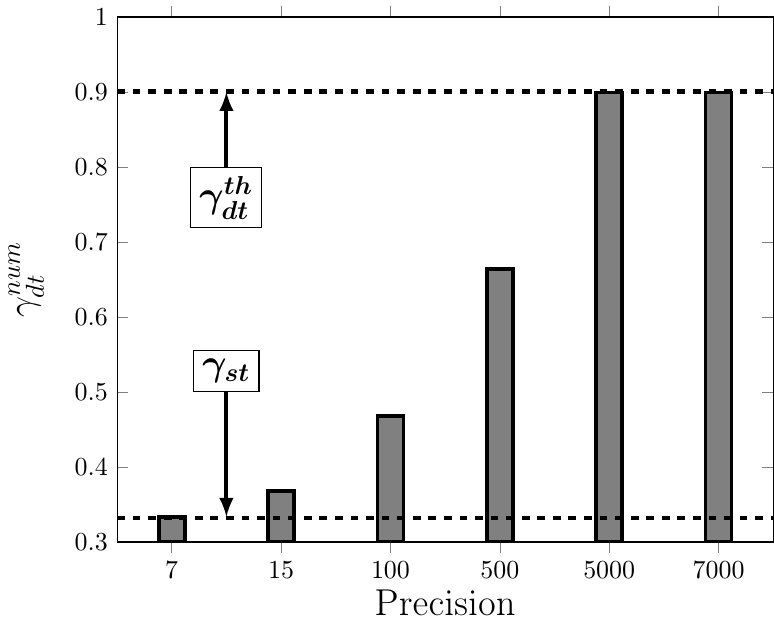}
\caption{Graphical representation of $\gamma_{dt}^{num}$ for different precisions (prec.~= 7, 15, 100, 500 and 5000) and for $\epsilon=10^{-4}$. Results are also compared to analytical \textit{static} and \textit{dynamic} thresholds: $\gamma_{st}$ and $\gamma_{dt}^{th}$. $\zeta=0.5$ and $\gamma_0 = 0$.}
\label{Fig_Influ_prec_1}
\end{figure}

\begin{figure*}[t]
\centering
\includegraphics[width=0.70\textwidth]{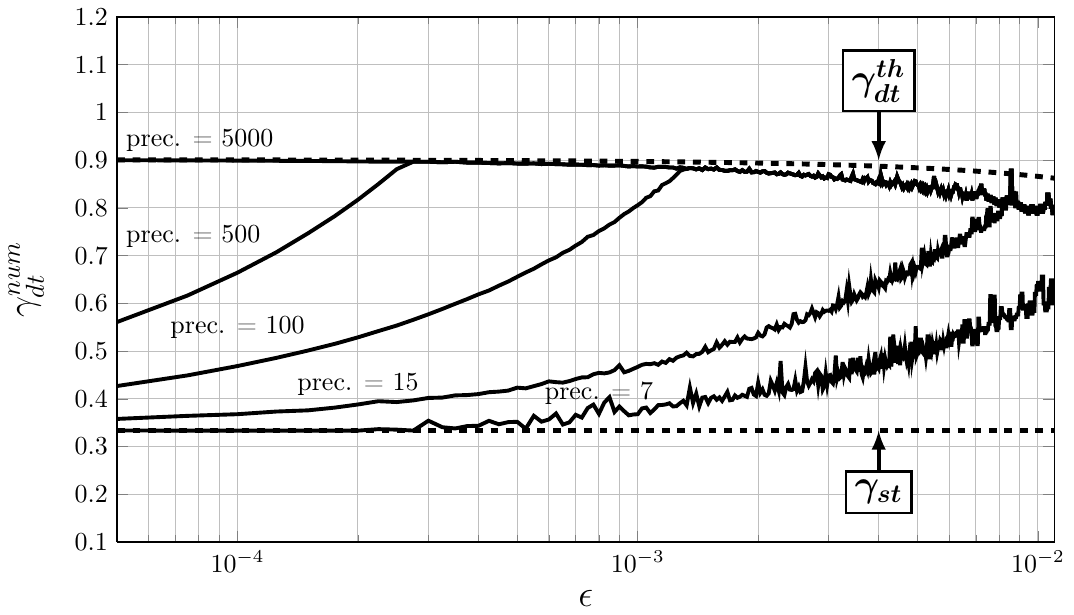}
\caption{Graphical representation of $\gamma_{dt}^{num}$ as a function of $\epsilon$ for $\zeta=0.5$, $\gamma_0=0$ and using five different precisions. A logarithmic scale is used in abscissa.}
\label{Fig_Influ_prec_2}
\end{figure*}

The first thing to observe in Fig.~\ref{Fig_Influ_prec_1} is the very high sensitivity of $\gamma_{dt}^{num}$ to precision, yet all the values of $\gamma_{dt}^{num}$ appear between $\gamma_{st}$ and $\gamma_{dt}^{th}$. For the lowest precision (7 decimal digits) the bifurcation delay disappears and $\gamma_{dt}^{num}=\gamma_{st}$. If the precision is very high (typically $\geq 5000$ decimals) $\gamma_{dt}^{num}=\gamma_{dt}^{th}$. Therefore, $\gamma_{dt}^{th}$ can be interpreted as the limit of the bifurcation delay when precision tends to infinity. In cases with intermediate precisions (prec.~= 15, 100 and 500) the bifurcation delay increases with the precision.

The sensitivity to the precision depends on the value of the increase rate $\epsilon$: Fig.~\ref{Fig_Influ_prec_2} plots $\gamma_{dt}^{num}$ with respect to $\epsilon$ for different values of the numerical precision. Results are also compared with $\gamma_{st}$ and $\gamma_{dt}^{th}$.

As above, for the lowest precision (7 decimals) the bifurcation delay disappears when $\epsilon$ is sufficiently small. Indeed, $\gamma_{dt}^{num}$ is constant and equal to $\gamma_{st}$. Then bifurcation delay occurs and increases with $\epsilon$. The case of the highest precision (5000 decimals) is identical to an analytical case which would correspond to infinite precision. When $\epsilon$ is sufficiently small, the curves of $\gamma_{dt}^{num}$ and $\gamma_{dt}^{th}$ overlap. In this case when $\epsilon$ is small $\gamma_{dt}^{num}$ is almost constant suggesting that the bifurcation delay does not depend on the increase rate, as previously shown in Fig.~\ref{seuildynVSci_d}. Then, still in the case of a precision of 5000 decimals, $\gamma_{dt}^{num}$ decreases for increasing $\epsilon$, and $\gamma_{dt}^{th}$ also decreases but to a lesser extent. For intermediate precisions (15, 100 and 500 decimals) the curve of $\gamma_{dt}^{num}$ first increases before stabilizing close to the curve of $\gamma_{dt}^{th}$.

For a given value of the precision, the larger   the $\epsilon$, the smaller is the accumulation of round-off errors created by the computer to reach a certain value of $\gamma$. This explains why the bifurcation delay first increases if the precision is not sufficiently high to simulate an analytic case. Beyond a certain value, all curves coincide with the one corresponding to the highest precision. That means that the system has reached the pair of parameters [precison ; $\epsilon$] needed to simulate an analytic case.

\section{Conclusion}

When considering mathematical models of musical instruments, oscillation threshold obtained through a static bifurcation analysis may be possibly very different from the threshold detected on a numerical simulation of this model. 
  
For the first time for musical instruments, the differences between these two thresholds have been interpreted as the appearance of the phenomenon of \textit{bifurcation delay} in connection with the concept of \textit{dynamic bifurcation}. 

Theoretical estimations of the dynamic bifurcation provided in this paper have to be compared with care to numerical simulations since the numerical precision used in computations plays a key role: for numerical  precisions close to standard machine precision, the bifurcation towards the oscillating regime can occur at significantly lower mouth pressure values (while different most of the time from the threshold obtained through static bifurcation theory). Moreover, in that case, the threshold at which the oscillations start becomes  more dependent on the increase rate of the mouth pressure.

The dependency on precision can be linked to the influence of noise generated by turbulence as the musician blows into the instrument. This would explain why the delays observed in artificially blown instruments are shorter than the predicted theoretical ones \cite{BBCFA2012Exp}. This will be the subject of further work on this subject, as well as the validity of these results for smoother curves of variation of the mouth pressure.

Moreover, in the light of results presented here for a basic model of wind instruments, varying the blowing pressure (even slowly) does not appear as the best way to experimentally determine Hopf bifurcations (static). In a musical context, since the blowing pressure varies through time, the dynamic threshold is likely to give more relevant informations than the static threshold, even if, in a real situation the influence of noise must be considered.

As a final remark, the simplistic model used in this work only describes one point per half-period of the sound played by the instrument. It is thus not suitable to describe different regimes (whose frequencies are harmonics of the fundamental one) that can be obtained by the instrument. However a simple extension of this model calculating the orbits of different instants within the half-period may be able to provide some insight on this subject.

\subsection*{acknowledgements}
We wish to thank Mr. Jean Kergomard for his valuable comments on the manuscript.

This work was done within the framework of the project SDNS-AIMV "Syst\`{e}mes Dynamiques Non-Stationnaires - Application aux Instruments \`{a} Vent" financed by \emph{Agence Nationale de la Recherche} (ANR).


\appendix
\gdef\thesection{Appendix \Alph{section}}
 
\section{Table of notation}
\label{App:A}

\subsection{Physical variables}

\begin{tabular}{p{1cm}p{4.6cm}p{1cm}}
\hline
\textbf{Symbol}   & \textbf{Explanation} & \textbf{Unit} \\ \hline
$Z_c$ & characteristic impedance  & Pa$\cdot$s$\cdot$m$^{-3}$ \\
$K_s$ & static stiffness of the reed per unit area & Pa$\cdot$m$^{-1}$ \\
$P_M$ & static closing pressure of the reed & Pa \\
$H$ & opening height of the reed channel at rest & m \\
$U$ & flow created by the pressure imbalance between the mouth and the mouthpiece & m$^3 \cdot$s$^{-1}$ \\
$U_r$ & flow created by the motion of the reed & m$^3 \cdot$s$^{-1}$ \\
$U_{in}$ & flow at the entrance of the resonator & m$^3 \cdot$s$^{-1}$ \\
$U_A$ & flow amplitude parameter & m$^3 \cdot$s$^{-1}$ \\
$P_m$ &  musician mouth pressure & Pa \\
$P$  & pressure inside the mouthpiece & Pa \\ 
$\Delta P$ & pressure difference $P_m-P$ & Pa \\  
$y$ & displacement of the tip of the reed & m \\   
$\tau$ & round trip travel time of a wave along the resonator & s \\
\hline
\end{tabular}

\subsection{Dimensionless variables}

\begin{tabular}{p{1cm}p{5.6cm}}
\hline
\textbf{Symbol}   & \textbf{Associated physical variable}  \\ \hline
$\gamma$ & musician mouth pressure \\
$\zeta$ & flow amplitude parameter \\
$u$ & flow at the entrance of the resonator \\
$p$  & pressure inside the mouthpiece \\
$r$ & reflexion function of the resonator \\
$p^+$ & outgoing wave \\
$p^-$ & incoming wave \\
$p^{+*}$ & non-oscillating static regime of $p^+$ (fixed points of the function $G$) \\
$\phi$ & invariant curve \\
$w$ & difference between $p^+$ and $\phi$ \\
$\epsilon$  &increase rate of the parameter $\gamma$ \\
$\gamma_{st}$ & static oscillation threshold \\
$\gamma_{dt}$ & dynamic oscillation threshold \\
$\gamma_{dt}^{th}$ & theoretical estimation of the dynamic oscillation threshold \\
$\gamma_{dt}^{num}$  & value of $\gamma$ when the system begins to oscillate (calculated numerically) \\
\hline
\end{tabular}

\subsection{Nonlinear characteristic of the embouchure}

\begin{tabular}{p{1.5cm}p{2.8cm}p{2cm}}
\hline
\textbf{Function}   & \textbf{Associated representation} & \textbf{Definition} \\ \hline
$F$ & $\{u \, ; \,p\}$ & $u=F(p)$ \\
$G$ & $\{p^+ \, ; \,p^-\}$ & $p^+=G(-p^-)$ \\
\hline
\end{tabular}

\section{Invariant curve}\label{an:incurve}

The invariant curve $\phi(\gamma,\epsilon)$ is invariant under the mapping (\ref{dynsys_pp}), it therefore satisfies the following equation:

\begin{equation}
\phi(\gamma,\epsilon)=G\left(\phi(\gamma-\epsilon,\epsilon),\gamma\right).
\label{eqdiff_1A}
\end{equation}

First of all, the invariant curve is expanded into a power series of $\epsilon$ and only he first-order is retained:

\begin{equation}
\phi(\gamma,\epsilon) \approx \phi_0(\gamma)+\epsilon \phi_1(\gamma).
\end{equation}

Secondly, the function $G$ is linearized around the curve $p^{+*}(\gamma)$ of the fixed points:

\begin{eqnarray}
G(x,\gamma) &\approx&G\left(p^{+*}(\gamma),\gamma\right)+  \nonumber \\ 
&&\left[x-p^{+*}(\gamma)\right]\partial_x G\left(p^{+*}(\gamma),\gamma\right) \\
 &=&p^{+*}(\gamma)+\left[x-p^{+*}(\gamma)\right]\partial_x G\left(p^{+*}(\gamma),\gamma\right),
\end{eqnarray}
where
\begin{equation}
\partial_x G\left(x,y\right)=\frac{\partial G(x,y)}{\partial x}.
\end{equation}

Then, we make a Taylor expansion of $\phi(\gamma-\epsilon,\epsilon)$:

\begin{eqnarray}
\phi(\gamma-\epsilon,\epsilon) &\approx& \phi(\gamma,\epsilon)-\epsilon\frac{\partial \phi}{\partial \gamma}(\gamma,\epsilon)+O(\epsilon^2);\\
&=& \phi_0(\gamma)+\epsilon \phi_1(\gamma) - \epsilon\frac{\partial \phi_0(\gamma)}{\partial \gamma}+O(\epsilon^2).
\end{eqnarray}

Finally, neglecting the second-order terms in $\epsilon$, equation (\ref{eqdiff_1A}) becomes:

\begin{multline}
\phi_0(\gamma)+\epsilon \phi_1(\gamma) = 
 p^{+*}(\gamma) \, +\\
\left[\phi_0(\gamma)+\epsilon \phi_1(\gamma) - \epsilon\frac{\partial \phi_0(\gamma)}{\partial \gamma}-p^{+*}(\gamma)\right] \times \\
 \partial_x G\left(p^{+*}(\gamma),\gamma\right).
\label{eqdiff_2A}
\end{multline}

To obtain the approximate analytical expression of the invariant cure $\phi$, equation (\ref{eqdiff_2A}) is successively solved for the functions $\phi_0(\gamma)$ and $\phi_1(\gamma)$.

As expected, to order 0 we find:

\begin{equation}
\phi_0(\gamma)= p^{+*}(\gamma).
\end{equation}

To order 1, we have to solve:

\begin{eqnarray}
\phi_1(\gamma) &=&
\left[\phi_1(\gamma) - \frac{\partial \phi_0(\gamma)}{\partial \gamma}\right]\partial_x G\left(p^{+*}(\gamma),\gamma\right);\\
&=& \left[\phi_1(\gamma) - \frac{\partial p^{+*}(\gamma)}{\partial \gamma}\right]\partial_x G\left(p^{+*}(\gamma),\gamma\right),
\end{eqnarray}
and therefore:

\begin{equation}
\phi_1(\gamma)=\frac{\partial p^{+*}(\gamma)}{\partial \gamma}
\frac{\partial_x G\left(p^{+*}(\gamma),\gamma\right)}{\partial_x G\left(p^{+*}(\gamma),\gamma\right)-1}.
\end{equation}

Finally the expression of the invariant curve is:

\begin{multline}
\phi(\gamma,\epsilon)\approx \\ p^{+*}(\gamma)\;+\;\epsilon\;\frac{\partial p^{+*}(\gamma)}{\partial \gamma}\;
\frac{\partial_x G\left(p^{+*}(\gamma),\gamma\right)}{\partial_x G\left(p^{+*}(\gamma),\gamma\right)-1}.
\end{multline}



\end{document}